\documentclass[%
 reprint,
 amsmath,amssymb,
 aps,
 prl,
]{revtex4-2}

\usepackage[utf8]{inputenc}
\usepackage{graphicx}
\usepackage{dcolumn}
\usepackage{bm}
\usepackage{braket}
\usepackage[dvipsnames]{xcolor}

\begin{document}

\preprint{APS/123-QED}

\title{Experimental investigation of superdiffusion via coherent disordered Quantum Walks}

\author{Andrea Geraldi$^1$}
\email{andrea.geraldi@uniroma1.it}
\author{Alessandro Laneve$^1$}
\author{Luis Diego Bonavena$^1$}
\author{Linda Sansoni$^1$}
\author{Jose Ferraz$^2$}
\author{Andrea Fratalocchi$^3$}
\author{Fabio Sciarrino$^1$}
\author{\\Alvaro Cuevas$^{1,4}$}
\author{Paolo Mataloni$^1$}
\affiliation{$^1$Dipartimento di Fisica, Sapienza Università di Roma, Piazzale Aldo Moro, 5, I-00185 Roma, Italy
}
\affiliation{$^2$Departamento de Física, Universidade Federal Rural de Pernambuco, 52171-900 Recife, Brazil
}
\affiliation{$^3$PRIMALIGHT, Faculty of Electrical Engineering; Applied Mathematics and Computational Science,
King Abdullah University of Science and Technology (KAUST), Thuwal 23955-6900, Saudi Arabia
}
\affiliation{$^4$ICFO-Institut de Ciències Fotòniques, The Barcelona Institute of Science and Technology, 08860 Castelldefels, Barcelona, Spain
}

\date{\today}

\begin{abstract}
Many disordered systems show a superdiffusive dynamics, intermediate between the diffusive one, typical of a classical stochastic process, and the so called ballistic behaviour, which is generally expected for the spreading in a quantum process.
We have experimentally investigated the superdiffusive behaviour of a quantum walk (QW), whose dynamics can be related to energy transport phenomena, with a resolution which is high enough to clearly distinguish between different disorder regimes.
By our experimental setup, the region between ballistic and diffusive spreading can be effectively scanned by suitably setting few degrees of freedom and without applying any decoherence to the QW evolution.
\end{abstract}

\maketitle


\textit{Introduction} - The study of energy transport phenomena in heterogeneous systems, like synthethic or biological media, ranging from tissues to computation nodes, enlightens the role of quantum coherence \cite{engel2007evidence,lambert2013quantum} which is believed to enhance the rate of these processes \cite{mohseni2008environment,hoyer2010limits}. In such conditions, a superdiffusive dynamic settles in the transport or propagation processes.
This is the case of heat excitation transport in particular condensed matter systems such as one dimensional (1D) metal lattices, described by the Luttinger liquids theory \cite{bulchandani2019superdiffusive}. Studies performed in this context, have shown a violation in the dichotomy between ballistic and diffusive transport regimes. There are several cases in which superdiffusion occurs, for example in classical one-dimensional systems \cite{van2012exact}, when both disorder and non linear effects are present \cite{kim2018direct}, and  even in quantum systems experiencing Many-Body Localization phase transitions \cite{vasseur2016nonequilibrium}.\\
Quantum walks (QWs) have long been found to efficiently describe coherent energy transport \cite{mohseni2008environment, hoyer2010limits}. 
By suitably introducing disorder in a QW, it is possible to modify its spreading behaviour. A paradigmatic example is given by Anderson localization \cite{anderson1958absence}, originally formulated for condensed matter systems, and extensively demonstrated in the case of all-optical systems \cite{segev2013anderson,schwartz2007transport,crespianderson2013}. The theoretical and experimental studies carried out on this topic have covered a wide range of phenomena, including "hyper-transport" of light \cite{levi2012hyper, schreiberecoherence2011}, and disordered QWs preserving the time dependence of spreading \cite{brun2003quantum}, and decoherence effects determining the transition from Quantum to Classical Random Walks (CRWs) \cite{schreiberecoherence2011, broomedecoherence2010}.
In these studies the role of disorder source was played by decoherence or various types of unitary evolutions.
The present work deals with the description of the transition between quantum and classical regimes, while the coherence of the evolution is preserved. This has been done experimentally by implementing proper configurations of disorder in a discrete Quantum Walk (QW).\\
\textit{Theoretical Model} - The theoretical description of a discrete QW is based on a walker, a coin and the operators acting on them \cite{venegasQW2012}\cite{kempeQW2003}. In a 1D QW, the walker is given by a quantum particle which is in a superposition of the position states, described by the set of states $\{\ket{i}_p\}$, each of them corresponding to a particular site of the line. The coin is represented by any internal binary degree of freedom of the particle, given by the basis kets $\{\ket{0}_c,\ket{1}_c\}$. The evolution of the quantum particle is controlled by the coin operator $\hat{C}$, acting on the coin state, and by the shift operator $\hat{S}$, which moves the walker according to the coin state. The evolution of the walker is then described by the repeated action of coin and shift operators on the general state of the particle $\ket{\psi}=\sum_{i,j}P_{i,j}\ket{i}_p\otimes\ket{j}_c$, $\ket{\psi'}=\hat{S}\cdot(\hat{I}\otimes \hat{C})\ket{\psi}$
where the coin operator is expressed as $\hat{C}=\frac{1}{\sqrt{2}}(\ket{0}_c\bra{0}_c+\ket{0}_c\bra{1}_c+\ket{1}_c\bra{0}_c-\ket{1}_c\bra{1}_c)$ while the shift operator reads
$\hat{S}=\sum_{i}\ket{i-1}_p\bra{i}_p\otimes\ket{0}_c\bra{0}_c+\ket{i+1}_p\bra{i}_p\otimes\ket{1}_c\bra{1}_c.$
These equations describe the evolution of a walker in a completely ordered QW, in which the coin operator is uniform both in space and time. Other structures of the coin operator, based on suitable maps of the QW phases, are possible \cite{ahlbrechtasymptotic2011}.\\
Here, by randomly shifting the phase difference imposed by the coin operator, we are able to simulate the superdiffusive behaviour of a quantum walker and to investigate the transition from the ballistic regime of an ordered QW to the diffusive one of a Classical Random Walk (CRW).\\
The operator describing the phase shift at a certain step $n$ of the evolution can be written as $ \hat{P}_{n}=\sum_{i}\ket{i}_p\bra{i}_p\otimes(\ket{0}_c\bra{0}_c+\ket{1}_c\bra{1}_ce^{i\Delta\phi^i(n)})$, where $\Delta\phi^i(n)$ corresponds to the phase difference imposed by the coin operator in the site $i$. We define the degree of disorder $p$ as the percentage of random phases that the walker experiences during the evolution. As a consequence, the case $p=0$ corresponds to a standard ordered QW while for $p=1$ the QW becomes completely disordered (case of a CRW). We call this kind of Quantum Walk $p$\emph{-diluted} QW, because of the dilution of a given degree of disorder $p$ during the QW evolution. With no further assumptions this model describes a completely space- and time-uncorrelated disorder.\\
We limit our analysis to the case in which the phases difference $\Delta\phi^i(n)$ can only be $0$ or $\pi$. We are interested to study how the quantum to classical transition depends on the parameter $p$. For a given value of $p$ a number of different phase maps can be realized, in principle, corresponding to different evolutions of the probability distribution. Thus, the values of the relevant parameters identifying a given value of $p$ are obtained by averaging over many different phase map configurations characterized by the same degree of disorder.\\
\begin{figure}[!h]
 	\centering
 	\includegraphics[scale=0.55]{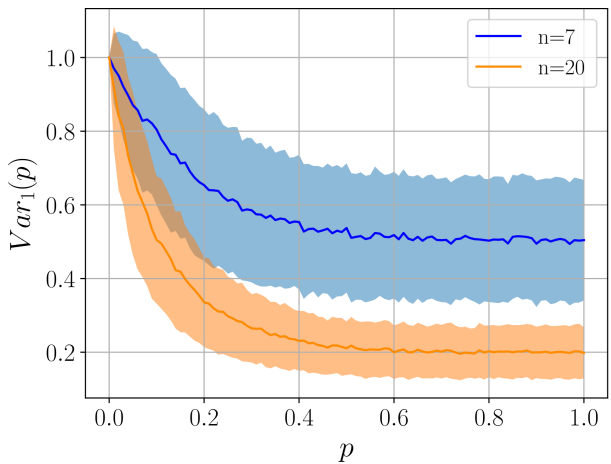}%
 	\caption{\emph{Simulation of the variance behaviour for 7 and 20 steps as a function of $p$. The average values of the variance have been calculated on a sample of 1000 different phase maps for each value of $p$. In the vertical axes $Var_1(p)$ for 7 and 20 steps have been normalized to their maximum values. Error bands correspond to the standard deviation of the variance distributions.}}
 	\label{fig:var_simulated_vs_p}
 \end{figure}
A useful quantity in the study of the single particle evolution is the variance of the position probability distribution, defined as:
\begin{equation}
 Var_{1}(n)=\sum_{i=-n}^ni^2\cdot P_i(n)-(\sum_{i=-n}^ni\cdot P_i(n))^2   
\end{equation}
where $P_{i}(n)$ is the probability to find the walker on site $i$ at step $n$.
We simulated the behaviour of the variance for different values of $p$ up to 20 steps, in the range $p=0$ to $1$, with a progressive increment of $0.01$. For each value of $p$ we simulated the evolution of $1000$ different phase maps, then we computed the mean value of the variance. In order to show how good this model of disorder is to study the superdiffusion, we computed the standard deviation of the variance distribution for each value of $p$. Results of simulation for 7 and 20 steps are shown in Fig. \ref{fig:var_simulated_vs_p}. In both cases the mean value of variance decreases with increasing of $p$, since a constant value around $p\sim 0.40$ is approached for both 7 and 20 steps. This indicates that the output distribution rapidly merges in a classical one for $p>0.40$, while a genuine superdiffusive behaviour occurs between $0<p\le 0.20$. Focusing on this region we observe that, for a given value of $Var_1(p)$, the uncertainty on the corresponding value of $p$ decreases with increasing of $n$.\\
A further useful quantity is the similarity which is a measure of how two distributions G and G' are similar: $S(G,G')=\frac{(\sum_{i}\sqrt{G_iG'_i})^2}{\sum_{i}G_i\sum_{j}G_j}$. In the graph of Fig. \ref{fig:KL_vs_p} we show the behaviour of the similarity between the average distribution for a given value of disorder $p$ and the ordered or disordered distribution at varying of $p$ (respectively, solid and dashed lines). The similarity with the ordered (disordered) distribution decreases (increases) with $p$ since the higher the disorder the more destructive interference tend to reproduce the disordered distribution. In the inset, the crossing points between the two quantities identify the transition between quantum and classical regimes for a given number of step $n$. It is worth noting that the transition point decreases with $n$. 
\begin{figure}[!h]
 	\centering
 	\includegraphics[scale=0.55]{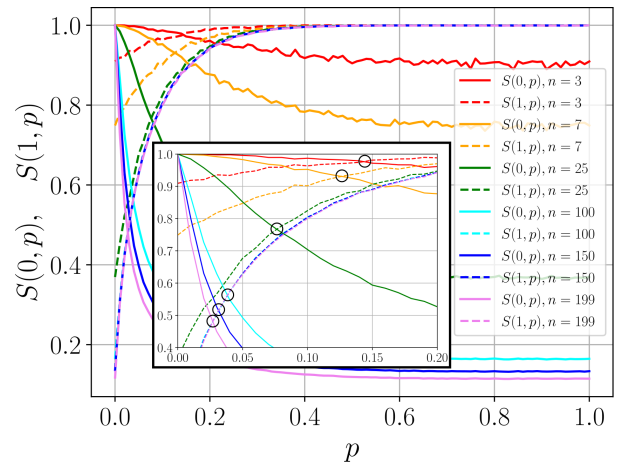}
 	\caption{\emph{Computational simulations for similarity between ordered (solid line) and disordered (dashed line) distributions (respectively $G(p=0)\equiv G(0)$ and $G(p=1)\equiv G(1)$) and the average distribution for a given value $p$ as a function of $p$. Here $S(0,p)$ $(S(1,p))$ stands for $S(G(0),G(p))$ $(S(G(1),G(p)))$. In the inset a zoom on the intersection region is reported. The crossing point shifts towards lower values of $p$ when the step number increases. Different colors stand for different step numbers of the evolution.}}
 	\label{fig:KL_vs_p}
 \end{figure}
This continuous transition occurs without loosing completely the quantum features of the evolution and can be interpreted as an evidence of the fact that the QW modifies its behaviour because of the total amount of disorder experienced along its spreading.\\
\textit{Experimental Implementation} - A 1D Quantum Walk can be realized by a network of Beam Splitters (BSs), each of them representing a particular site of the line and acting both as $\hat{C}$ and $\hat{S}$ operators. Several implementations of optical QW have been realized, based on bulk optics schemes, bulk-fiber circuits and femtosecond laser written photonic circuits \cite{bulkQW2005, crespianderson2013, silberhornQW2016, flaminireview2018, sansoni2012two}.
The bulk optic setup used in the present work consists of two displaced multi-pass Sagnac Interferometers ($SIs$) linked by a common Beam Splitter ($BS_1$ in Fig. \ref{fig:setup}).

\begin{figure}[!h]
 	\centering
 	\includegraphics[scale=0.23]{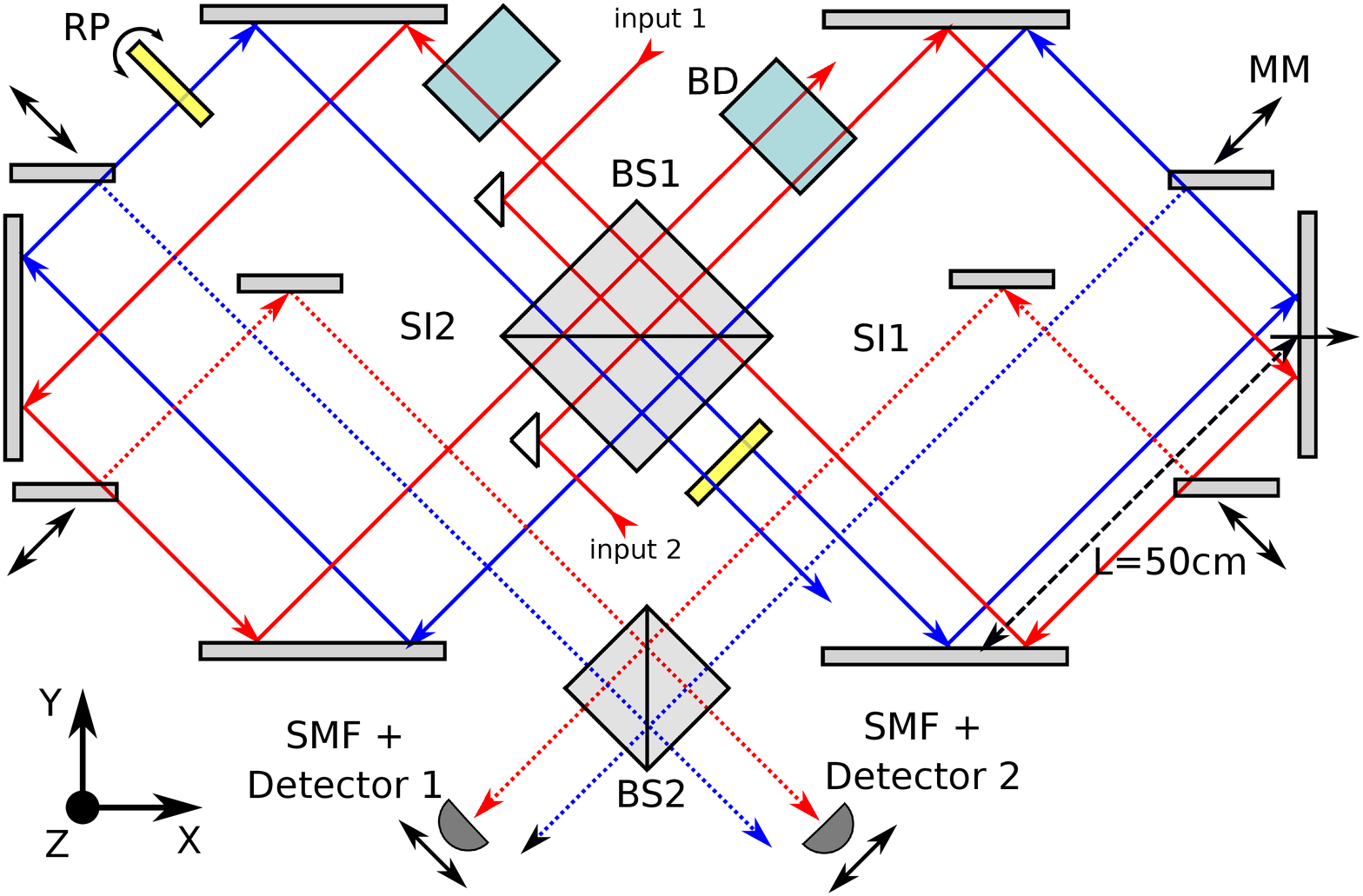}%
 	\caption{\emph{Sketch of the setup used in the experiment. BS: beam splitter, BD: beam displacer, RP: rotating glass plates, MM: moving mirror, SMF: single mode fiber. Each side of the square $SI$ is $50$cm long. Blue and red beams circulate in opposite directions and impinge on the $BS_1$ in the same horizontal point but at different heights along the z direction, due to the effect of BD.}}
 	\label{fig:setup}
 \end{figure}

\begin{figure}[!h]
 	\centering
 	\includegraphics[scale=0.60]{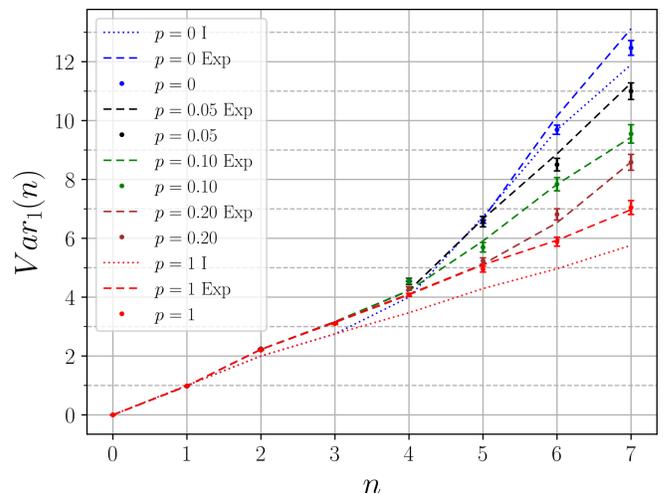}%
 	\caption{\emph{Experimental values of the variance as a function of the number of steps for different values of $p$ ($0,0.05,0.1,0.2,1$). Dots correspond to experimental data. For a given value of $p$, dashed line shows the expected average behaviour (Exp) of variance for the three selected phase maps. They were computed by taking into account the actual optical parameters of the setup. Dotted lines show the ideal behaviour (I) of a perfectly symmetric $BS_1$ in the two limit cases, namely the ordered and disordered QW. Note that, up to the third step, the variance is not affected by the phase thus the red and blue dotted lines are coincident; the same effect occurs for all the dashed lines.}}
 	\label{fig:var_vs_step_1photon}
 \end{figure}

Beam displacement is obtained by translating one mirror of $SI_1$ along the $x$ direction. This particular configuration allows to create in the $x-y$ horizontal plane of Fig. \ref{fig:setup} a (quasi) infinite loop, which is equivalent to a chain of phase-stable and independently tunable Mach-Zehnder interferometers \cite{cuevasmarkovianity2019}. The entire BS network necessary for a QW is realized by exploiting also the $z$ direction. On this purpose, suitable Beam Displacers (BDs) are inserted along clockwise directions in $SI_1$ and counterclockwise ones in $SI_2$ allowing to increase the number of $x-y$ planes on which the photons can travel (see Supplementary Material (SM)). Phases are independently addressed in each unit mesh of the QW by using Rotating Glass Plates (RP). The output modes of each step can be extracted for measurement by a set of Moving Mirros (MMs), intersecting and extracting from the setup only the modes at the selected step $N_j$, without intersecting the steps $N<N_j$. The extracted radiation is then coupled on a single mode fiber and measured (for further details on the setup see \cite{cuevasmarkovianity2019} and \cite{geraldiQW2019}). A slight modification of the setup allows to inject two distinguishable or indistinguishable photons through the two different input ports of $BS_1$. After the extraction of the modes, a further BS ($BS_2$) allows to separate photons travelling along the same mode. Adding a further coupler intercepting the extracted modes, coincidences between all possible modes at a given step can be measured and the whole two photon probability distribution can be experimentally reconstructed (see SM for further details).\\
\textit{Experimental Results} - It is well known that in a superdiffusive process the spread of the walker position follows a power law, expressed in our case by $Var_1(n)\propto n^{\beta}$, with $n$ corresponding to the number of steps and $1<\beta<2$ \cite{havlin2002diffusion}. Here, the case $\beta=2$ ($\beta=1$) is typical of a QW (CRW).
We experimentally implemented five different values of $p$, namely $0, 0.05, 0.10, 0.20$ and $1$. For each value of $p>0$ we selected three different phase maps and experimentally reconstructed the output probability distribution for each step of the evolution. In Fig. \ref{fig:var_vs_step_1photon} the average variance of the walker is reported as a function of the step number $n$.

\begin{figure}[!h]
 	\centering
 	\includegraphics[scale=0.60]{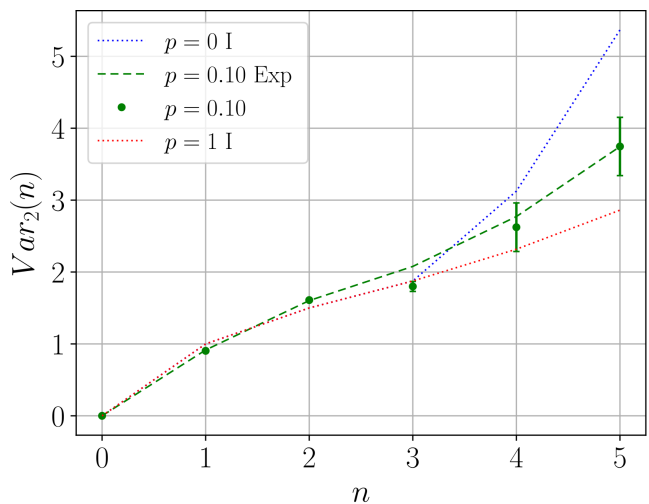}%
 	\caption{\emph{Experimental results for the two photon QW. Dots represent experimental data, dashed lines indicate the expected behaviour obtained by taking into account the real parameters of the setup (Exp), dotted lines correspond to the limit cases of the ideal ordered and disordered two photon QW (I). Note that, up to the third step, the variance is not affected by the phase thus the red and blue dotted lines are coincident.}}
 	\label{fig:var_vs_step_2photons}
 \end{figure}

The experimental setup allows to start the QW only from the central position of the 1D-lattice and by using any of the two coin modes (input ports of $BS_1$). Accordingly, we studied the evolution for the input state $\ket{\psi}=\ket{0}_{p}\otimes\ket{0}_{c}$.
Experimental data are in excellent agreement with the theoretical prediction for the three chosen phase maps (dashed lines). The whole superdiffusive region, namely the one between the diffusive (red lines) and the ballistic (blue lines) behaviours, can be exploited by varying the values of $p$. A small amount of phase instability was present due to the large size of the apparatus. This effect has been included in the computation of errors bars. We also show in Fig. \ref{fig:var_vs_step_1photon} the expected behaviours simulated by considering ideal optical elements for the two limit cases of ordered and completely disordered QW (respectively blue and red dotted lines). They differ from the experimental data because of the non perfectly symmetric behaviour of $BS_1$ (see SM for details) \cite{geraldiQW2019}.\\
The actual superdiffusive behaviour of the experimental evolutions is shown in Tab. \ref{tab:fit_results}, where we report the values of $\beta$ obtained by fitting the experimental data ($\beta_{Fit}$) and comparing them with the values obtained through theoretical simulations ($\beta_{Theo}$). The observed discrepancies can be explained as follows:

\begin{itemize}
    \item The $\beta_{Theo}$ values have been obtained without considering the experimental components imperfections, e. g., BS asymmetry and losses.
    \item Numerical simulations for $p>0$ have been obtained, as said, by averaging over $1000$ disorder configurations, while three phase maps were used in the experiment. Thus, we expect the simulation results to be far more accurate, while the experimental ones are more sensitive to stochastic deviations.
\end{itemize}
It is worth noting that, for the ordered case, the values of $\beta_{Theo}$ and $\beta_{Fit}$ are lower than the expected value of $2$. Indeed the quadratic growth is an asymptotic behaviour, expected to be achieved for long evolution times.
We can conclude that a superdiffusive behaviour, uniquely identified by the condition $1<\beta<2$, can be simulated using $p$-diluted QWs, making them a useful tool for further investigations on diffusion processes.
 \begin{center}
 \begin{table}[!h]
 \color{black}{
 {\small{}
  \hfill{}
  \begin{tabular}{c|c|c}
  \em $p$ & \em $\beta_{Theo}$ & \em $\beta_{Fit}$ \\
  \hline
  $0$ & 1.69 & 1.64 $\pm$ 0.10 \\
  \hline
  $0.05$ & 1.540 & 1.433 $\pm$ 0.067 \\
  \hline
  $0.10$  & 1.414 & 1.277 $\pm$ 0.061 \\
  \hline
  $0.20$ & 1.198 & 1.160 $\pm$ 0.060 \\
  \hline
  $1$ & 0.921 & 0.961 $\pm$ 0.022
 \end{tabular}}
 }
 \hfill{}
  \caption{\emph{Expected ($\beta_{Theo}$) and experimental ($\beta_{Fit}$) values of the exponent $\beta$ for different values of $p$. The values between $1$ and $2$ for $p>0$ confirm the superdiffusive behaviour of the evolution.}}
 \label{tab:fit_results}
 \end{table}
 \end{center}
We also investigated the case of two photon QW, with photon pairs entering into the setup through the two $BS_1$'s input ports (see Fig.3). The variance of the mean position of two indistinguishable walkers, whose expression is
\begin{equation}
\label{vartwophotons}
Var_{2}(n)=\sum_{i,j=-n}^n(\frac{i+j}{2})^2\cdot P_{i,j}(n)-(\sum_{i,j=-n}^n\frac{i+j}{2}\cdot P_{i,j}(n))^2
\end{equation}
is given in Fig. \ref{fig:var_vs_step_2photons} for $p=0.10$. In this case the number of steps is limited to $n=5$ mostly for two reasons: the losses present in the apparatus, limiting the number of detectable coincidences, and the growth of the number of modes on which the photons can travel. Experimental data are in very good agreement with theoretical simulation taking into account the real setup parameters. Simulations of the behaviour of $Var_2$ in the ordered and disordered case are also reported. Error bars are larger than the ones in the single photon case due to the fact that in this case a little variation of the phase must be considered twice, because it is experienced by both photons (see SM for further details on the two photon case).

\textit{Conclusion} - We have experimentally investigated the superdiffusion process, intermediate between the quantum and classical transport regimes, by introducing tunable disorder, through suitable phase map configurations, in a discrete QW. The intrinsically stable bulk optic system used in the experiment allows to operate with any kind of phase map. By increasing the number of steps exploited by the optical setup, the subdiffusive regime, localized between the diffusive and the localized one, can also be experimentally explored.
The $p$-diluted QW approach described in this work makes possible to simulate a decoherent process without altering the coherent evolution of the walker.
We are currently investigating on the relationship existing between the two models. 

\date{\today}

\begin{acknowledgments}
We acknowledge support from the European Commission grants FP7-ICT-2011-9-600838 (QWAD - Quantum Waveguides Application and Development). We thank Mauro Paternostro, Beatrice Polacchi, Camilla Sarra and Federico Pegoraro for useful help and discussions.
\end{acknowledgments}

\bibliography{ms}

\newpage

\section{Supplementary Material to\\Experimental investigation of superdiffusion via coherent disordered Quantum Walks}
\section{Experimental setup}
For a given step $n$ of the QW there are $n+1$ possible sites on which the walker can be found and $2n$ output optical modes. Half of them represents the state $\ket{0}$ of the coin (the ones circulating in the counterclockwise direction in $SI_1$ and in the clockwise one in $SI_2$) while the remaining modes correspond to the $\ket{1}$ state of the coin (the ones circulating in the clockwise direction in $SI_1$ and in the counterclockwise one in $SI_2$). By using the set of MMs (see Fig. 3 of the main text), each mode can be independetly extracted and measured.\\
The setup allows to start the QW with any input state by injecting photons in a linear superposition of the two coin states, corresponding to the input ports of $BS_1$.\\
Being a closed loop, the system allows in principle a very large number of steps. However this is limited mainly by photon beam divergence. By reducing both the total length of the optical setup and the transverse distance of the travelling beams, we expect to double the maximum achievable number of steps.\\
\begin{figure}[!h]
	\includegraphics[scale = 0.30]{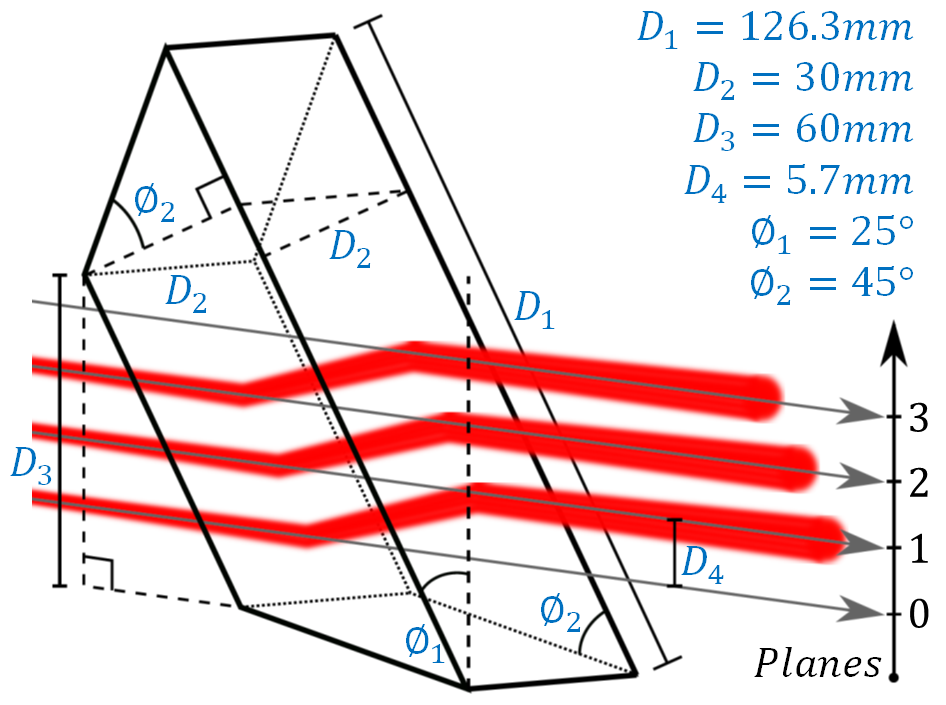}
	\caption{\emph{Sketch of the functioning of the BD used in the actual setup. Values of relevant geometrical parameters are reported.
	}}
	\label{fig:BD}
\end{figure}
The measured reflectivity value of both $BS_1$ and $BS_2$ used in the experiment was $R=0.45$. This asymmetry determines, together with other effects, mainly caused by the transmission losses of the optical elements of the setup, a behaviour of the actual QW which is discordant from the ideal one. This has been taken into account in simulations described in the main text.\\
The Beam Displacer (BD) consists of a thick inclined glass prism. Its operation is sketched in Fig. \ref{fig:BD}. When a beam passes through it, it is vertically shifted by a fixed amount ($\sim 5$mm), determined by inclination, thickness and refractive index of the BD glass. A critical parameter is given by the parallelism between the BD's faces, $<1\mu$rad in our case. This guarantees an interference visibility larger than $95\%$ up to the seventh step, mostly limited by unavoidable experimental errors in the setup alignment.

\section{Experimental results: Single photon}
\begin{figure*}[!h]
	\includegraphics[scale = 0.49]{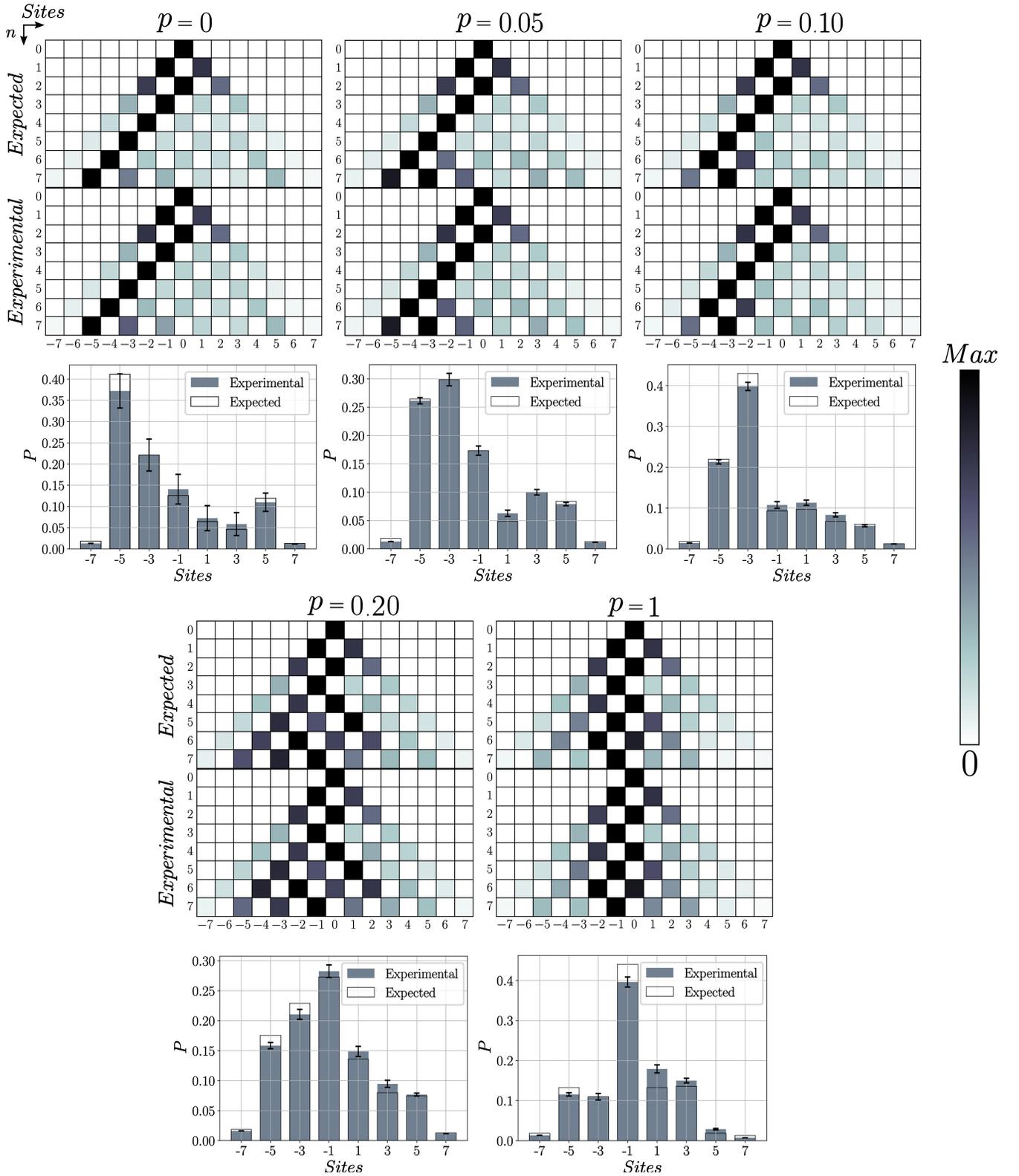}
	\caption{\emph{Density plots: evolution of the probability distributions for $p=0,0.05,0.10,0.20,1$ in the single photon case. Experimental data and simulations taking into account real parameters of the optical setup are shown. Bar plots: seventh step probability distributions for different values of $p$. Black-contoured bars represent expected values taking into account real parameters of the setup, gray bars represent experimental data. Data for $p>0$ are the result of the average over three different phase maps.
	}}
	\label{fig:evolutions}
\end{figure*}

In Fig. \ref{fig:evolutions} we show the density plots of the evolutions of the single photon QW in the cases $p=0, 0.05, 0.10, 0.20$ and $1$. The horizontal axis represents different sites that the walker can reach while the vertical one corresponds to the step number, increasing from top to bottom. For the sake of clarity, data for each step are normalized to the maximum value. For each value of $p>0$ we averaged the probability distributions over three experimental realizations with different phase maps. Both experimental data and theoretical predictions are reported by taking into account the real parameters of the optical setup. High similarity values between theoretical and experimental distributions are obtained for each value of $p$, as shown in Tab. \ref{tab:similarities_single_photon}. The maximum of the probability distribution is localized around site $-5$ for the ordered case which is typical for a walker propagating in an ordered QW \cite{venegasQW2012}. For larger values of $p$ ($0.05, 0.10$) the maximum shifts towards higher site numbers ($-3$). For $p=0.20$ the mean position of the distribution approaches the starting position $0$ and, in the case of complete disorder, the walker tends to reproduce the CRW distribution. This behaviour can more easily observed in the bar plots distribution corresponding to the seventh step given for the different values of $p$. Black-contoured bars represent theoretical predictions by taking into account real parameters of the optical setup while gray bars correspond to the experimental data.\\

\begin{center}
	\begin{table}[!h]
		\color{black}{
			{\small{}
				\hfill{}
				\begin{tabular}{c|c}
					\em $p$ & \em Similarity \\
					\hline
					$0$ & 0.999 $\pm$ 0.001 \\
					\hline
					$0.05$ & 0.999 $\pm$ 0.001 \\
					\hline
					$0.10$  & 0.998 $\pm$ 0.002 \\
					\hline
					$0.20$ & 0.998 $\pm$ 0.002 \\
					\hline
					$1$ & 0.993 $\pm$ 0.001
			\end{tabular}}
		}
		\hfill{}
		\caption{\emph{Similarities between expected probability distributions and experimental ones for the single photon QW for different values of $p$. Errors computation takes into account a small amount of phase instability due to large dimensions of the optical setup.}}
		\label{tab:similarities_single_photon}
	\end{table}
\end{center}

\section{Experimental results: two photons}
\begin{figure}[!h]
	\centering
	\includegraphics[scale=0.31]{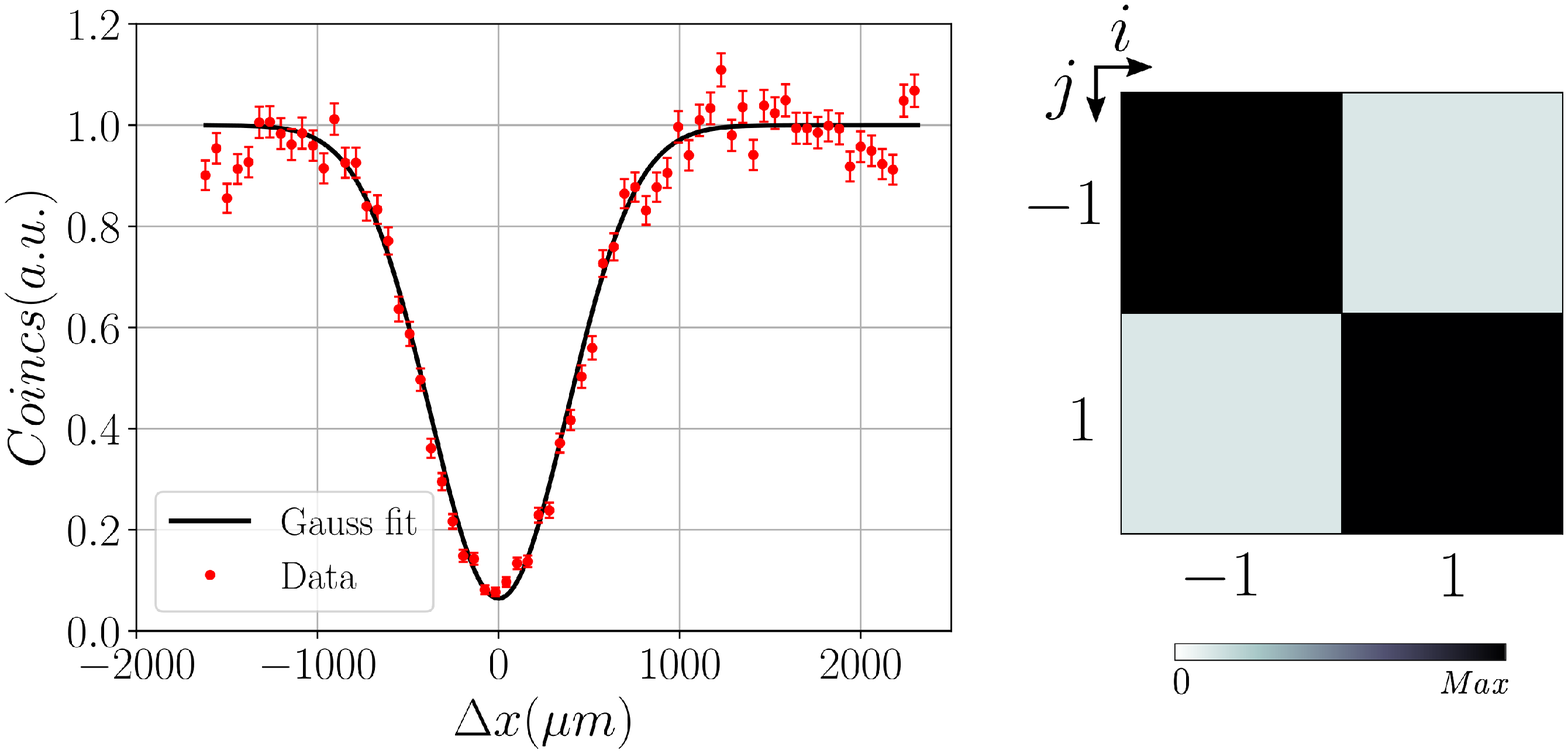}
	\caption{\emph{Left: normalized coincidences vs photons optical delay. Red dots represent experimental data with error bars computed considering poissonian distributed coincidences. Black line represents a gaussian fit of the experimental data. Photon indistinguishability is confirmed by the high dip visibility $V\sim 93\%$. Right: normalized experimental coincidences matrix for the first step of the QW. $i$ ($j$) represents sites of the first (second) detector.}}
	\label{fig:HOM_dip}
\end{figure}

In order to inject two indistinguishable photons, synchronization of their arrival time on the $BS_1$ was necessary. To this aim, a Hong-Ou-Mandel dip has been measured (see Fig.\ref{fig:HOM_dip}).
\begin{center}
	\begin{table}[!h]
		\color{black}{
			{\small{}
				\hfill{}
				\begin{tabular}{c|c}
					\em $n$ & \em Similarity \\
					\hline
					$1$ & 0.999 $\pm$ 0.001 \\
					\hline
					$2$ & 0.999 $\pm$ 0.001 \\
					\hline
					$3$  & 0.99 $\pm$ 0.02 \\
					\hline
					$4$ & 0.99 $\pm$ 0.05 \\
					\hline
					$5$ & 0.97 $\pm$ 0.05
			\end{tabular}}
		}
		\hfill{}
		\caption{\emph{Similarities between expected coincidence matrices and experimental ones for the two photons QW up to the fifth step. Errors computation takes into account a small amount of phase instability due to large dimensions of the optical setup.}}
		\label{tab:similarities_two_photons}
	\end{table}
\end{center}

\begin{figure*}
	\centering
	\includegraphics[scale=0.30]{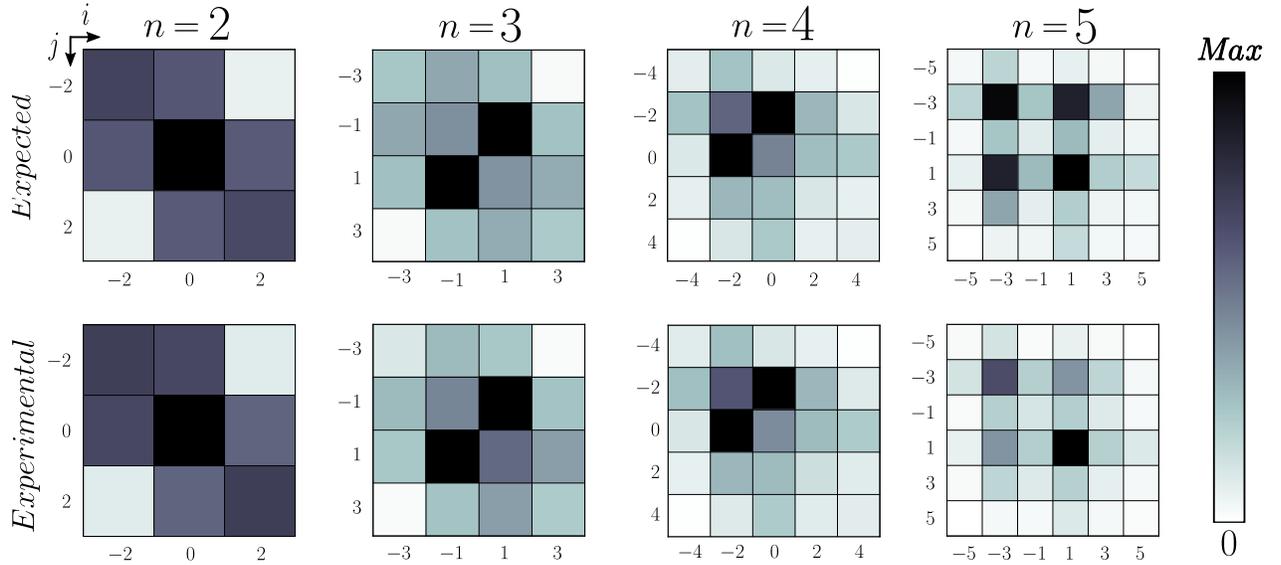}
	\caption{\emph{Probability distributions up to the fifth step when two indistinguishable photons enter the QW. For a better clarity each matrix is normalized to its own maximum value. Similarities between expected and experimental matrices are reported in Tab.\ref{tab:similarities_two_photons}.}}
	\label{fig:coinc_matrices}
\end{figure*}

With accidental coincidence correction, the measured visibility was $V\sim 93\%$. The condition of equal arrival time on the $BS_1$ corresponds to the first step of the QW with indistinguishable photons, whose coincidences matrix is reported on the right of Fig. \ref{fig:HOM_dip}. In order to measure the probability to find two photons travelling along the same output mode, a further BS ($BS_2$) was used to separate them (see Fig. 3 of the main text). In this way it was possible to measure coincidences between all possible pairs of modes. For the first step we observe a minimum probability to measure a coincidence between two different modes and a maximum probability to find two photons travelling the same mode.
We reconstructed the coincidence matrices up to the fifth step in the case of $p=0.10$. Experimental data, together with theoretical predictions are shown in Fig. \ref{fig:coinc_matrices}. Similarities between experimental and expected results are high, demonstrating the feasibility of the apparatus even for two photon measurements (see Tab. \ref{tab:similarities_two_photons}).

\end{document}